\documentclass[reprint, double column, superscriptaddress,prl, showkeys]{revtex4-1}

\usepackage{float}
\usepackage{graphicx,epsfig}
\usepackage{amssymb}
\usepackage{amsmath}
\usepackage{bm}
\usepackage{braket}

\usepackage{textcomp}
\usepackage{color}
\usepackage{pgffor}

\usepackage{soul}

\usepackage[bookmarksnumbered,bookmarksopen]{hyperref}
\hypersetup{
   colorlinks=true,       
}

\usepackage{pdfpages}
\usepackage{BOONDOX-cal}

\makeatletter

\AtBeginDocument{\let\LS@rot\@undefined}
\makeatother

\newif\ifarXiv
\arXivtrue

\usepackage{soul}

\begin{document}
\setcounter{page}{1}

\title[]{Fractional Quantum Hall State at Filling Factor $\nu=1/4$ in Ultra-High-Quality GaAs 2D Hole Systems}
\author{Chengyu \surname{Wang}}
\author{A. \surname{Gupta}}
\author{S. K. \surname{Singh}}
\author{P. T. \surname{Madathil}}
\author{Y. J. \surname{Chung}}
\author{L. N. \surname{Pfeiffer}}
\author{K. W. \surname{Baldwin}}
\affiliation{Department of Electrical and Computer Engineering, Princeton University, Princeton, New Jersey 08544, USA}
\author{R. \surname{Winkler}}
\affiliation{Department of Physics, Northern Illinois University, DeKalb, Illinois 60115, USA}
\author{M. \surname{Shayegan}}
\affiliation{Department of Electrical and Computer Engineering, Princeton University, Princeton, New Jersey 08544, USA}
\date{\today}

\begin{abstract}
Single-component fractional quantum Hall states (FQHSs) at even-denominator filling factors may host non-Abelian quasiparticles that are considered to be building blocks of topological quantum computers. Such states, however, are rarely observed in the lowest-energy Landau level, namely at filling factors $\nu<1$. Here we report evidence for an even-denominator FQHS at $\nu=1/4$ in ultra-high-quality two-dimensional hole systems confined to modulation-doped GaAs quantum wells. We observe a deep minimum in the longitudinal resistance at $\nu=1/4$, superimposed on a highly insulating background, suggesting a close competition between the $\nu=1/4$ FQHS and the magnetic-field-induced, pinned Wigner solid states. Our experimental observations are consistent with the very recent theoretical calculations which predict that substantial Landau level mixing, caused by the large hole effective mass, can induce composite fermion pairing and lead to a non-Abelian FQHS at $\nu=1/4$. Our results demonstrate that Landau level mixing can provide a very potent means for tuning the interaction between composite fermions and creating new non-Abelian FQHSs.
\end{abstract}

\maketitle

Even-denominator fractional quantum Hall states (FQHSs) are fascinating condensed matter phases. The best-known example is the even-denominator FQHS at  Landau level (LL) filling factor $\nu=$5/2 observed in GaAs two-dimensional electron systems (2DESs) when a first excited ($N=1$) spin LL is half-occupied \cite{Willett.PRL.1987, Pan.PRL.1999}. It is generally believed to be a BCS-type, paired state of flux-particle composite fermions (CFs) \cite{Jain.PRL.1989, Moore.NPB.1991, Greiter.PRL.1991, Read.PRB.2000}. This state may have non-Abelian quasiparticles as its excitations, and be of potential use in fault-tolerant, topological quantum computing \cite{Nayak.RMP.2008, Banerjee.Nature.2018, Willett.PRX.2023}.

The CF pairing that leads to the stability of the $\nu=$5/2 FQHS is facilitated by the node in the in-plane wavefunction of electrons in the $N=1$ LL as it allows them to come closer to each other. Such pairing is much harder to achieve in the ground state ($N=0$) LL, consistent with the near absence of even-denominator FQHSs. Instead, the ground state at $\nu=$1/2 (and 1/4) is a compressible CF Fermi sea, flanked by a plethora of odd-denominator FQHSs at nearby fillings \cite{Jain.Book.2007}. An exception is a 2DES with \textit{bilayer} charge distribution. A FQHS at $\nu=$1/2 was observed in 2DESs confined to wide GaAs quantum wells (QWs) \cite{Suen.PRL.1992, Suen.PRL.1992B, Suen.PRL.1994, Shabani.PRB.2013} and double QWs \cite{Eisenstein.PRL.1992}.  These were originally interpreted as a two-component, Abelian FQHS described by the Halperin-Laughlin ($\psi_{331}$) wavefunction \cite{He.PRB.1993, Halperin.1983}, with the layer or electric-subband index playing the role of an extra degree of freedom. Although the two-component origin of the $\nu=1/2$ FQHS is widely accepted for the double QWs where interlayer tunneling is negligible, recent experiments \cite{Mueed.PRL.2015, Mueed.PRL.2016, Singh.preprint.2023} and theory \cite{Zhu.PRB.2016} suggest that in wide QWs where interlayer tunneling is significant, the $\nu=1/2$ FQHS is likely a single-component, non-Abelian state. In addition, another even-denominator FQHS was reported in wide GaAs QWs at $\nu=$1/4 \cite{Luhman.PRL.2008, Shabani.PRL.2009}, and theory suggests it is also likely a single-component, non-Abelian state, topologically equivalent to an $f$-wave paired state of CFs \cite{Faugno.PRL.2019}. We emphasize that, for both $\nu=$1/2 and 1/4 FQHSs in wide QWs, the thick and bilayer-like charge distribution is crucial as it leads to a softening of the Coulomb repulsion and CF pairing. 

\begin{figure*}[t!]
  \begin{center}
    \psfig{file=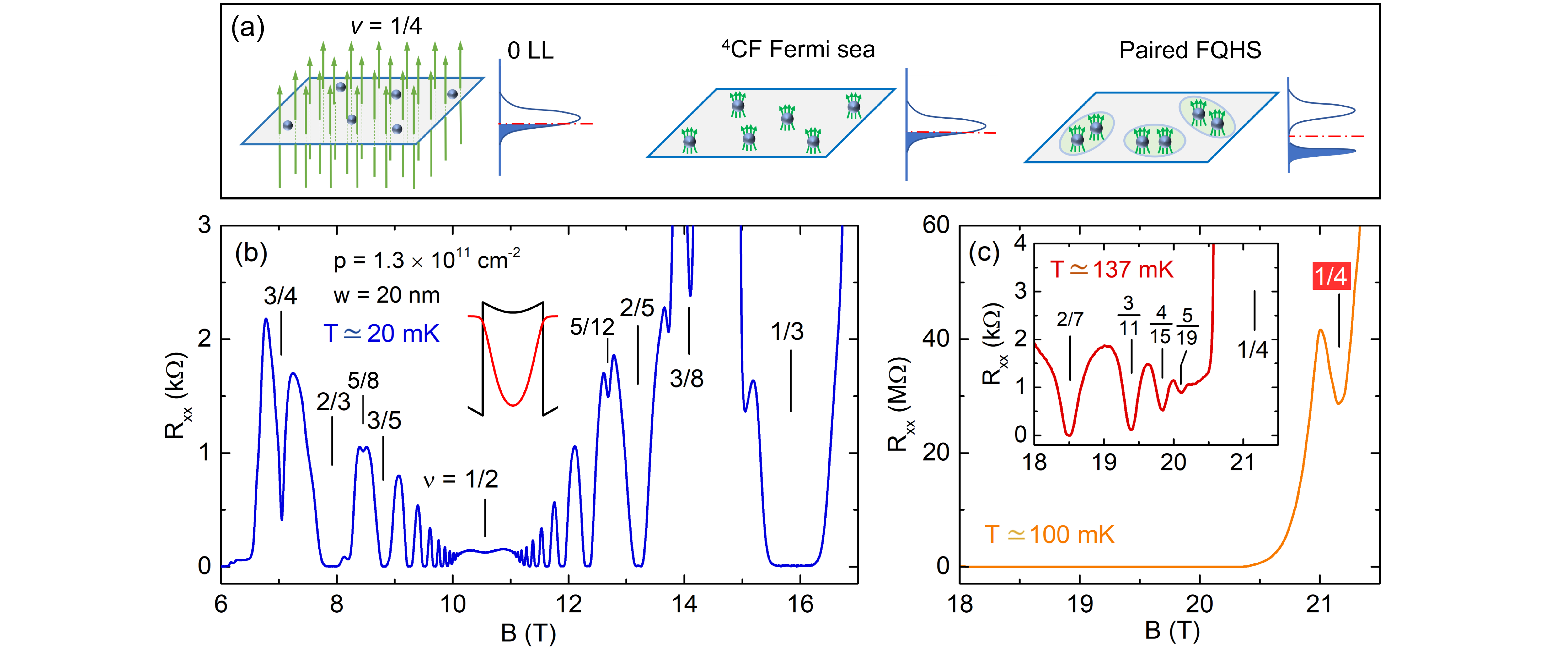, width=1\textwidth}
  \end{center}
  \caption{\label{test}
    (a) Schematics of the pairing mechanism for even-denominator FQHS at $\nu=$1/4. The blue spheres represent holes, and the green vertical arrows the magnetic field flux quanta. The curved short arrows in the right two panels represent magnetic flux quanta attached to holes to form four-flux composite fermions ($^4$CFs). If LLM is strong, $^4$CFs pair and condense to form a FQHS at $\nu=$1/4. (b) $R_{xx}$ vs $B$ trace near $\nu=$1/2 for a 2DHS, with density $1.3\times10^{11}$ cm$^{-2}$ and QW width 20 nm, taken at $T\simeq20$ mK with $I=$20 nA. Inset shows the self-consistently calculated hole charge distribution (red) and potential (black). (c) $R_{xx}$ vs $B$ trace near $\nu=$1/4, taken at $T\simeq$100 mK with $I=$0.1 nA. Inset shows $R_{xx}$ vs $B$ between $\nu=$2/7 and 1/4 taken at $T\simeq$137 mK. A higher current ($I=$50 nA) is used to reduce noise.
    }
  \label{fig:test}
\end{figure*}

Here, we report experimental evidence for a developing FQHS at $\nu=$1/4 in ultra-high-quality 2D \textit{hole} systems (2DHSs) confined to \textit{narrow} GaAs QWs with \textit{single-layer} charge distributions. We attribute this surprising observation to the much larger effective mass of GaAs 2D holes ($m^*\simeq$0.5, in units of free electron mass) \cite{Zhu.SSC.2007} compared to electrons ($m^*=$0.067), and the ensuing severe LL mixing (LLM). LLM is often parameterized as the ratio of the Coulomb energy to cyclotron energy, $\kappa=(e^2/4\pi \epsilon l_B) / (\hbar eB/m^*)$, and is proportional to $m^* B^{1/2}$, where $l_B = (\hbar /eB)^{1/2}$ is the magnetic length. LLM can play a crucial role in determining the many-body ground states in different 2D material systems including semiconductor heterostructures \cite{Rezayi.PRL.2017, Das.PRL.2023} and atomically-thin 2D materials (e.g., monolayer graphene \cite{Peterson.PRL.2014}). For example, it can affect the stabilization of possible non-Abelian FQHSs at $\nu=$5/2 \cite{Rezayi.PRL.2017, Das.PRL.2023} and high-field Wigner crystal \cite{Zhao.PRL.2018}. Most relevant to our study, very recent theoretical calculations suggest that substantial LLM can destabilize the CF Fermi sea at even-denominator fillings in the lowest LL and lead to the emergence of a single-component, non-Abelian FQHS through CF pairing [Fig. 1(a)] \cite{Zhao.PRL.2023}. Our results, together with the calculations of Ref. \cite{Zhao.PRL.2023}, establish ultra-high-quality GaAs 2DHSs as a platform for hosting and creating exotic, non-Abelian FQHSs through LLM.

We studied 2DHSs in GaAs QWs grown on GaAs (001) substrates using molecular beam epitaxy. The samples have Al$_x$Ga$_{1-x}$As barriers and are modulation doped with C. They were grown following the optimization of the growth chamber vacuum integrity and the purity of the source materials \cite{Chung.NM.2021}, and have extremely high mobilities, up to $\simeq 6\times10^6$ cm$^2$/Vs \cite{Chung.PRM.2022}. The widths ($w$) of the QWs range from 20 to 35 nm, and their 2D hole densities ($p$) from 0.41 to 1.3, in units of $10^{11}$ cm$^{-2}$, which we use throughout this manuscript. We performed our experiments on $4\times 4$ mm$^2$, van der Pauw samples, with alloyed In:Zn contacts at the four corners and side midpoints. We cooled the samples in a dilution refrigerator, and measured the longitudinal resistance ($R_{xx}$) using the conventional lock-in amplifier techniques.

In Figs. 1(b,c) we present $R_{xx}$ vs magnetic field ($B$) traces for a 2DHS with $p=$1.3 and $w=$20 nm. Near $\nu=$1/2, the sample exhibits a smooth and shallow minimum, flanked by numerous odd-denominator FQHSs following the Jain sequence $\nu=n/(2n\pm1)$, where $n$ is an integer \cite{Jain.PRL.1989, Jain.Book.2007}. This is consistent with a compressible Fermi sea of two-flux CFs ($^2$CFs) being the ground state when the lowest spin LL is half-occupied \cite{Jain.Book.2007}. We also observe emerging even-denominator FQHSs at $\nu=$3/4, 3/8, 5/8, and 5/12 \cite{Wang.PRL.2022, Wang.PNAS.2023}. In Fig. 1(c) and its inset, we show $R_{xx}$ at higher $B$ at elevated temperatures ($\simeq$100 and 137 mK). For $18<B<20.5$ T, $R_{xx}$ remains in the k$\Omega$ range, and we observe odd-denominator FQHSs at $\nu=$2/7, 3/11, 4/15, and 5/19. These are the Jain-sequence states of 4-flux CFs ($^4$CFs) that follow $\nu=n/(4n-1)$ \cite{Jain.Book.2007}. Their presence, together with the higher-order FQHSs flanking $\nu=$1/2, attest to the exceptionally high quality of the 2DHS. 

In Fig. 1(c), when $B$ exceeds 20.5 T, $R_{xx}$ sharply increases and attains values $\simeq$40 M$\Omega$, even at a relatively high temperature of $\simeq$100 mK. The 2DHS becomes highly insulating in this field range, as we demonstrate later in this Letter. Such $B$-induced insulating phases have been previously reported in GaAs 2DESs at $\nu\lesssim1/5$ \cite{Jiang.PRL.1990, Goldman.PRL.1990, Deng.PRL.2018} and in GaAs 2DHSs at $\nu\lesssim1/3$ \cite{Santos.PRL.1992, Santos.PRB.1992, Ma.PRL.2020}. They are generally believed to signal the formation of Wigner solids (WSs) pinned by the ubiquitous disorder \cite{Santos.PRL.1992, Santos.PRB.1992, Ma.PRL.2020, Zhao.PRL.2018}.

The highlight of our study is the observation, for the first time, of a very deep and sharp $R_{xx}$ minimum at $\nu=$1/4, signaling a developing FQHS at this filling. The fact that the $R_{xx}$ minimum at $\nu=$1/4 appears on top of the insulating background suggests a close competition between the FQHS and WS states at $\nu=$1/4. This is reminiscent of the recent observation of a developing FQHS at $\nu=$1/7 in ultra-high-mobility 2DESs, also competing with surrounding WS states \cite{Chung.PRL.2022, Footnote.Hall, SM}.

\begin{figure*}
  \begin{center}
    \psfig{file=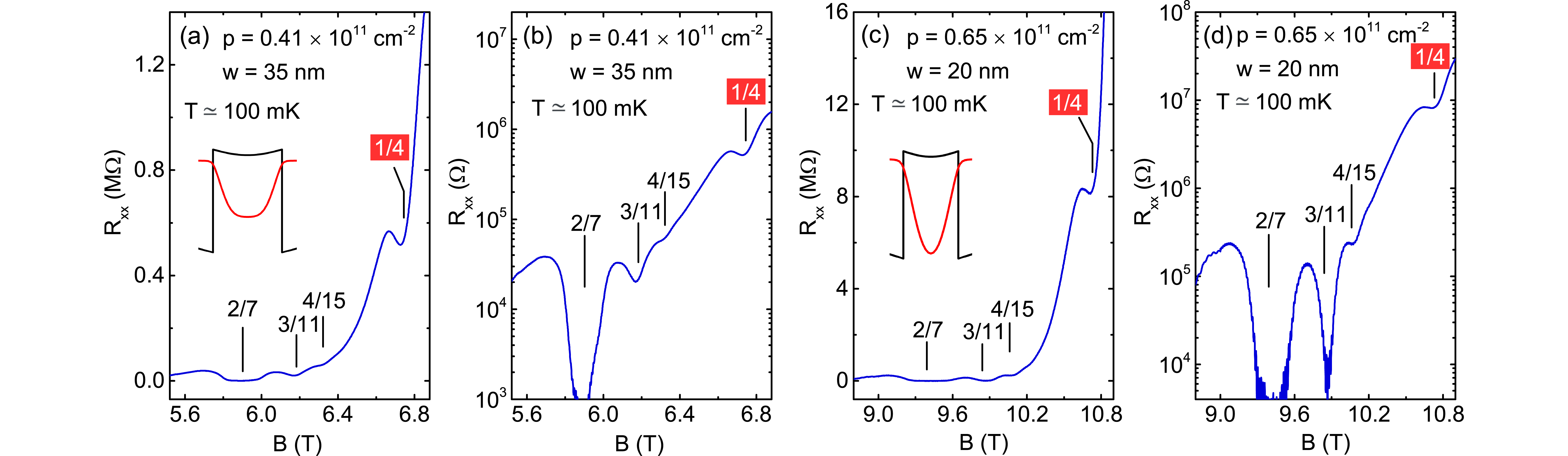, width=1\textwidth}
  \end{center}
  \caption{\label{samples} 
     $\nu=$1/4 FQHSs in 2D hole samples with different densities: (a,b) $p=0.41\times10^{11}$ cm$^{-2}$, and (c,d) $p=0.65\times10^{11}$ cm$^{-2}$. (b) and (d) are replots of (a) and (c) in $log$ scales for $R_{xx}$. Insets in (a) and (c): Self-consistently calculated hole charge distribution (red) and potential (black).
}
  \label{fig:samples}
\end{figure*}

\begin{figure}[t!]
  \begin{center}
    \psfig{file=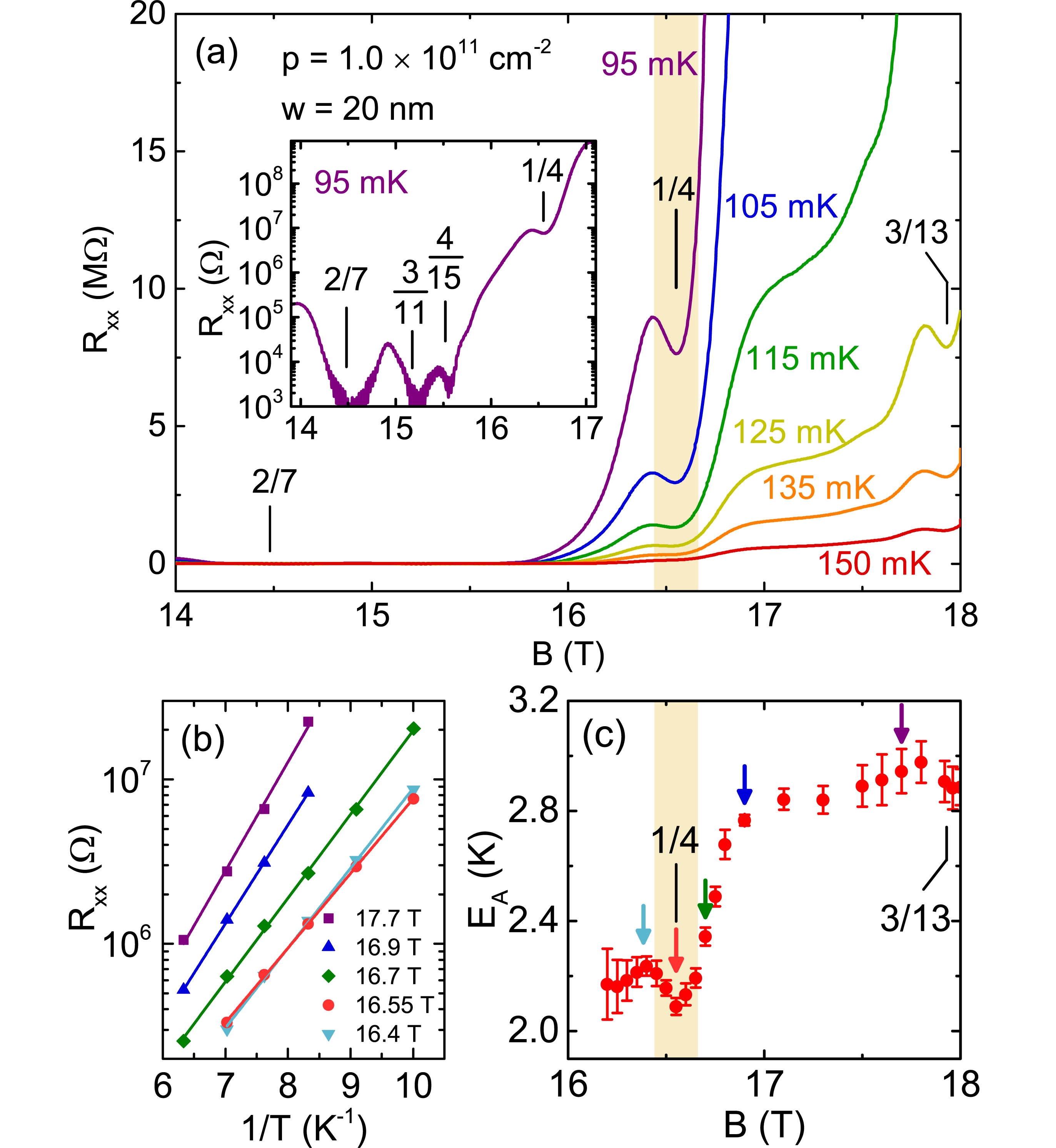, width=0.5 \textwidth}
  \end{center}
  \caption{\label{Tdep} 
   (a) $R_{xx}$ vs $B$ traces taken at different temperatures for a 2DHS with $p=1.0\times10^{11}$ cm$^{-2}$. Inset: $R_{xx}$ in $log$ scale vs $B$ at $T\simeq95$ mK. (b) Arrhenius plots of $R_{xx}$ vs $1/T$ at various magnetic fields, color-coded according to the fields marked by arrows in (c). The solid lines are linear fits to the data points according to $R_{xx}\propto e^{E_A/2kT}$, from which we obtain the activation energy $E_A$. (c) $E_A$ vs $B$.
  }
  \label{fig:Tdep}
\end{figure}

In order to confirm that the $R_{xx}$ minimum we observe at $\nu=$1/4 is intrinsic to ultra-high-quality 2DHSs, we measured several samples from different wafers with various hole densities. In Figs. 2 and 3, we present data for three samples with $p=0.41$, 0.65, and 1.0, and QW widths 35, 20, and 20 nm, respectively. Similar behavior is observed in all samples. On the low-field side of $\nu=1/4$, we observe $R_{xx}$ minima at odd-denominator $\nu=$2/7 and 3/11, and a minimum or an inflection point at 4/15, signaling developing FQHSs belonging to the $\nu=n/(4n-1)$ sequence; these are seen more clearly in $log$ scale plots of Figs. 2(b,d) and Fig. 3(a) inset. As we increase $B$ and approach $\nu=$1/4, $R_{xx}$ grows very rapidly, and the 2DHSs enter the insulating phase. Remarkably, in all samples, a well-defined and sharp $R_{xx}$ minimum is seen at $\nu=$1/4 superimposed on the insulating background.

We also investigated the temperature dependence of the $\nu=$1/4 FQHS and its adjacent insulating phases. Figure 3(a) illustrates $R_{xx}$ traces at various temperatures for a sample with $p=$1.0, and $w=$20 nm. As $T$ increases, the $R_{xx}$ minimum at $\nu=$1/4 gradually weakens and turns into an inflection point at 150 mK, while the background resistance decreases by more than ten times. At $B \simeq 17.9$ T and intermediate temperatures, we observe a developing FQHS at $\nu=$3/13, following the Jain-sequence states of $^4$CF [$\nu=n/(4n+1)$].

To quantitatively analyze the insulating phases near $\nu=$1/4, we deduce the activation energy $E_A$ from the relation $R_{xx}\propto e^{E_A/2kT}$. The energy scale $E_A$ is generally associated with the defect formation energy of the WS \cite{Archer.PRB.2014, Chung.PRL.2022}. In Fig. 3(b), we show the Arrhenius plots of $R_{xx}$ vs $1/T$. The slopes of the linear fits yield $E_A$ at a given $B$. In Fig. 3(c), we present $E_A$ for $16<B<18$ T, where $R_{xx}$ exhibits insulating behavior. The data reveal a clear minimum in $E_A$ at $\nu=$1/4 and a small dip at $\nu=$3/13. Similar features were reported in GaAs 2DESs at very low fillings, e.g. at $\nu=$1/7 and 2/13, in recent measurements of ultra-high-mobility 2DES samples \cite{Chung.PRL.2022}, and were interpreted as precursors to developing FQHSs \cite{Footnote.precursor.FQHS, Mendez.PRB.1983, Chang.PRL.1984, Willett.PRB.1988, Goldman.PRL.1993, Sajoto.PRL.1993}. We also note that the magnitude of $E_A$ in Fig. 3(c), $\simeq2$-3 K, is comparable to $E_A$ measured at similar $B$ near $\nu=$1/7 \cite{Chung.PRL.2022}. It is consistent with the theoretically calculated energies for the quantum bubble defect formation in WSs composed of CFs \cite{Archer.PRB.2014}, and suggests that the insulating behavior adjacent $\nu=$1/4 is intrinsic and originates from many-body phenomena rather than single-particle Anderson localization \cite{Footnote.insulating}.

An important requirement for a FQHS to be non-Abelian is that it is fully spin polarized \cite{Nayak.RMP.2008, Hossain.PRL.2018}. Our observation of the $\nu=$1/4 FQHS in a wide range of large perpendicular magnetic fields ($7\lesssim B_{\perp}\lesssim 21$ T) in samples with different densities strongly suggests that it is fully spin polarized. This conclusion is supported by our data taken in tilted magnetic fields, which reveal that the $\nu=$1/4 $R_{xx}$ minimum is always present, even when a large parallel magnetic field of $\simeq12$ T is applied (see Supplemental Material (SM) \cite{SM}). 

To understand the origin of the observed $\nu=$1/4 FQHS in our 2DHSs, it is helpful to compare our results to other 2D systems. In GaAs 2DESs confined to narrow QWs, the ground state at $\nu=$1/4 is a Fermi sea of $^4$CFs. This is supported by transport experiments where $R_{xx}$ is featureless near $\nu=$1/4, and odd-denominator Jain-sequence states of $^4$CF at $\nu=n/(4n\pm 1)$ are observed \cite{Pan.PRB.2000, Chung.PRL.2022}. The Fermi wavevector of the $^4$CF Fermi sea has indeed been directly measured by geometrical resonance \cite{Hossain.PRB.2019}. In 2DESs confined to wide GaAs QWs, however, a FQHS was observed at $\nu=1/4$ \cite{Luhman.PRL.2008, Shabani.PRL.2009}. This is explained by theory, suggesting that the large electron layer thickness in wide QWs can modify the interaction, making the CF Fermi sea at $\nu=$1/4 unstable to $f$-wave pairing \cite{Faugno.PRL.2019}. We can easily rule out the above mechanism for the $\nu=$1/4 FQHS in our 2DHSs because our samples are all narrow QWs, and have single-layer charge distributions (see insets to Figs. 1 and 2).

A $\nu=1/4$ FQHS has also been reported at an isospin transition in monolayer graphene \cite{Zibrov.NatPhys.2018}. This was interpreted as a multicomponent FQHS incorporating correlations between electrons on different carbon sublattices \cite{Zibrov.NatPhys.2018}. Similar physics was also observed in GaAs 2D holes: When the two lowest-energy LLs cross at $\nu=1/2$, an even-denominator FQHS manifests itself as the ground state \cite{Liu.PRB.2014}. However, it is extremely unlikely that the $\nu=1/4$ FQHSs in our 2D hole samples originate from a LL crossing. The LL crossing at $\nu=1/2$ or other fillings in 2DHSs occurs for carefully-tuned sample parameters (density, QW width, tilt angle, etc.) \cite{Liu.PRB.2014, Liu.PRB.2016, Ma.PRL.2022, Wang.PRL.2023}. In contrast, we observe a $\nu=1/4$ FQHS in several samples spanning a wide range of densities, QW widths, and tilt angles. It is hard to imagine that a crossing of two LLs occurs at $\nu=1/4$ in all these samples.

Recent theoretical calculations by Zhao \textit{et al}. \cite{Zhao.PRL.2023} offer a potential explanation for our experimental findings. Using fixed-phase diffusion Monte Carlo calculations, the authors predict that significant LLM could give rise to the pairing of CFs at 1/4 filling, leading to the destabilization of CF Fermi sea and the emergence of non-Abelian FQHSs [Fig. 1(a)] \cite{Zhao.PRL.2023}. Specifically, they propose a transition from a compressible CF Fermi sea to an incompressible, non-Abelian, paired FQHS \cite{Footnote.pairing} at $\nu=1/4$ when the LLM parameter $\kappa$ reaches a value of $\simeq 6$ to 7. This critical value is much higher than $\kappa$ for GaAs 2DESs, typically $\lesssim1$, consistent with the Fermi sea as the CF ground state at $\nu=1/4$.

GaAs 2D holes, however, have a much larger $\kappa$ because of their larger effective mass. For the samples investigated in our study, if we use the calculated $B=0$ effective mass and simply assume linear LLs as a function of $B$, we estimate that $\kappa$ at $\nu=1/4$ is between 3 and 8, close to the critical $\kappa$ estimated by the calculations \cite{Zhao.PRL.2023}. However, this agreement is likely fortuitous, because both the $\kappa$ we quote for our samples and the calculated critical $\kappa$ are rough estimates \cite{Footnote.estimate, SM, Zhao.PRL.2023}. 

A natural question that arises is the relation between the $\nu=1/4$ FQHS and the FQHS at $\nu=3/4$ reported recently in 2DHSs with similar ultra-high quality \cite{Wang.PRL.2022}. Assuming particle-hole symmetry in the lowest spin LL, these FQHSs may be viewed as a pair of particle-hole conjugate states. While severe LLM is evidently crucial in stabilizing both the 3/4 and 1/4 FQHSs, LLM is also known to break the $\nu \leftrightarrow (1-\nu)$ particle-hole symmetry, suggesting that the two states likely have distinct origins. This conjecture is supported by the very different behaviors for GaAs 2DHSs near $\nu=1/4$ and 3/4. On the flanks of $\nu=1/4$, we observe numerous odd-denominator FQHSs at $\nu=n/(4n-1)$. In contrast, no signs of odd-denominator FQHSs are observed at $\nu=1-n/(4n-1)$ in the same samples (see Fig. S3 in SM \cite{SM}). Furthermore, the theory of Ref. \cite{Zhao.PRL.2023} for $\nu=1/4$ cannot explain other even-denominator FQHSs observed in the lowest LL of 2DHSs, e.g., at $\nu=$3/8, 5/8, and 5/12. On the other hand, instead of being viewed as the particle-hole counterpart of the 1/4 FQHS, the FQHS at 3/4 filling of the 2DHS LL can be mapped to an \textit{effective filling} $\nu^*=3/2$ of $^2$CF Lambda levels \cite{Wang.PRL.2022}. In this picture, the excited CF Lambda level is half-occupied, reminiscent of the 5/2 FQHS in GaAs 2DESs where the excited electron spin LL is half-occupied. Similar explanations can be applied to the FQHSs at $\nu=$3/8, 5/8, and 5/12 \cite{Wang.PNAS.2023}. 

In closing, we emphasize that we observe signatures of the 1/4 FQHS in a highly-resistive, insulating regime where there is a close competition between the FQHSs at fractional fillings and the (pinned) WS states at nearby fillings. The observation of insulating phases is consistent with previous experiments and theories which have shown that severe LLM in 2DHSs can favor WS states at $\nu\lesssim 1/3$ \cite{Santos.PRL.1992, Santos.PRB.1992, Ma.PRL.2020, Zhao.PRL.2018}. The 1/4 FQHS therefore manifests itself amidst a challenging landscape where a small but finite amount of disorder can disturb the many-body ground states. The significant improvement in the quality of GaAs 2DHSs \cite{Chung.PRM.2022} is evidently playing a crucial role in the unraveling of intriguing interaction phenomena previously concealed because of the presence of higher amounts of disorder. While our report of new correlated states in an ultra-pure GaAs 2DHS advances the field of many-body condensed matter physics, it should also stimulate related future work in other 2D carrier systems, including atomically-thin 2D materials such as transition-metal dichalcogenides (e.g., WSe$_2$) and bilayer graphene. These materials also experience severe LLM in the quantum Hall regime \cite{Shi.NatNano.2020, Huang.PRX.2022} and, if their quality could be further improved, they should provide a fertile ground for the observation of exotic even-denominator FQHSs in the lowest LL.

\begin{acknowledgments}

We acknowledge support by the U.S. Department of Energy (DOE) Basic Energy Sciences (Grant No. DEFG02-00-ER45841) for measurements, the National Science Foundation (NSF) (Grant No. DMR 2104771 and No. ECCS 1906253) for sample characterization, and the Eric and Wendy Schmidt Transformative Technology Fund and the Gordon and Betty Moore Foundation’s EPiQS Initiative (Grant No. GBMF9615 to L.N.P.) for sample fabrication. Our measurements were partly performed at the National High Magnetic Field Laboratory (NHMFL), which is supported by the NSF Cooperative Agreement No. DMR 2128556, by the State of Florida, and by the DOE. This research is funded in part by QuantEmX grant from Institute for Complex Adaptive Matter and the Gordon and Betty Moore Foundation through Grant No. GBMF9616 to C. W., A. G., P. T. M., S. K. S., and M. S. We thank A. Bangura, R. Nowell, G. Jones, and T. Murphy at NHMFL for technical assistance, and A. C. Balram and J. K. Jain for illuminating discussions.

\end{acknowledgments}



\section*{Supplementary material}

# Supplemental Material to "Fractional Quantum Hall State at Filling Factor $\nu = 1/4$ in Ultra-High-Quality GaAs 2D Hole Systems"

Chengyu Wang,[1] A. Gupta,[1] S. K. Singh,[1] P. T. Madathil,[1] Y. J. Chung,[1] L. N. Pfeiffer,[1] K. W. Baldwin,[1] R. Winkler,[2] and M. Shayegan[1]

[1]*Department of Electrical and Computer Engineering,*
*Princeton University, Princeton, New Jersey 08544, USA*

[2]*Department of Physics, Northern Illinois University, DeKalb, Illinois 60115, USA*

(Dated: November 13, 2023)

# I. LANDAU LEVEL DIAGRAMS FOR GaAs TWO-DIMENSIONAL HOLE SYSTEMS

In Fig. S1, we present calculated energy ($E$) vs magnetic field ($B$) Landau level (LL) diagrams for our GaAs two-dimensional hole systems (2DHSs) with different hole densities $p$ and well width $w$. In each plot, the dash-dotted red curve traces the Fermi energy, and the $B$ position of $\nu = 1/2$ and $1/4$ are marked. The calculations were performed using the multi-band envelope function approximation based on the $8 \times 8$ Kane Hamiltonian [1]. The model also uses axial approximation which neglects the effect of cubic anisotropy of the crystal structure. Because of the significant heavy-hole light-hole mixing and spin-orbit coupling, the LLs for GaAs 2DHSs are highly nonlinear as a function of magnetic field $B$ and exhibit numerous crossings. For example, in Fig. S1(a), there are two crossings between the two lowest-energy LLs; the first one occurs at $\sim 2$ T and the second one at $\sim 18$ T. (By lowest-energy we mean LLs with the least negative values of $E$. In other words, the two lowest-energy LLs are the two top-most LLs at a particular $B$.) $\nu = 1/2$ is located between the first and second crossings while $\nu = 1/4$ is located on the higher-field side of the second crossing. The LLs shown in Figs. S1(b, c) are qualitatively similar except that the positions of the crossings are shifted and $\nu = 1/4$ is now also located between the first and second crossings. The calculated LLs for the 35-nm-wide 2DHS [Fig. S1(d)] is quite different from the other three: The LLs are overall closer in energy, suggesting a smaller cyclotron energy at a given $B$. In addition to the first and second crossings between the two lowest-energy LLs, there is a third crossing occurring at $\sim 28$ T.

LL mixing is parameterized by $\kappa$ which is defined as the ratio between Coulomb energy ($E_{Coul}$) and cyclotron energy ($E_{cyc}$). For GaAs 2D electrons whose LLs are linear as a function of $B$, $E_{Coul} \propto \sqrt{B}$ and $E_{cyc} \propto B$. As a result, $\kappa \propto 1/\sqrt{B}$, and therefore, $\kappa$ increases monotonically with decreasing density. The non-linearity of hole LLs adds complexity to the quantitative evaluation of LL mixing for GaAs 2DHSs as there is no well-defined cyclotron energy in magnetic field. As shown in Table I and discussed in detail later, it is very difficult to make good estimations of $\kappa$ for GaAs 2DHSs. Although we do not have a quantitative evaluation of $\kappa$, it is evident that $E_{Coul}$ is much larger than $E_{cyc}$ near $\nu = 1/2$ and $1/4$. For example, $E_{Coul}$ at 10 T is $\simeq 14$ meV. As seen in Fig. S1, there are numerous LLs between 0 and -6 meV at 10 T for all the four samples, indicating that $\kappa$ is much larger than 1.

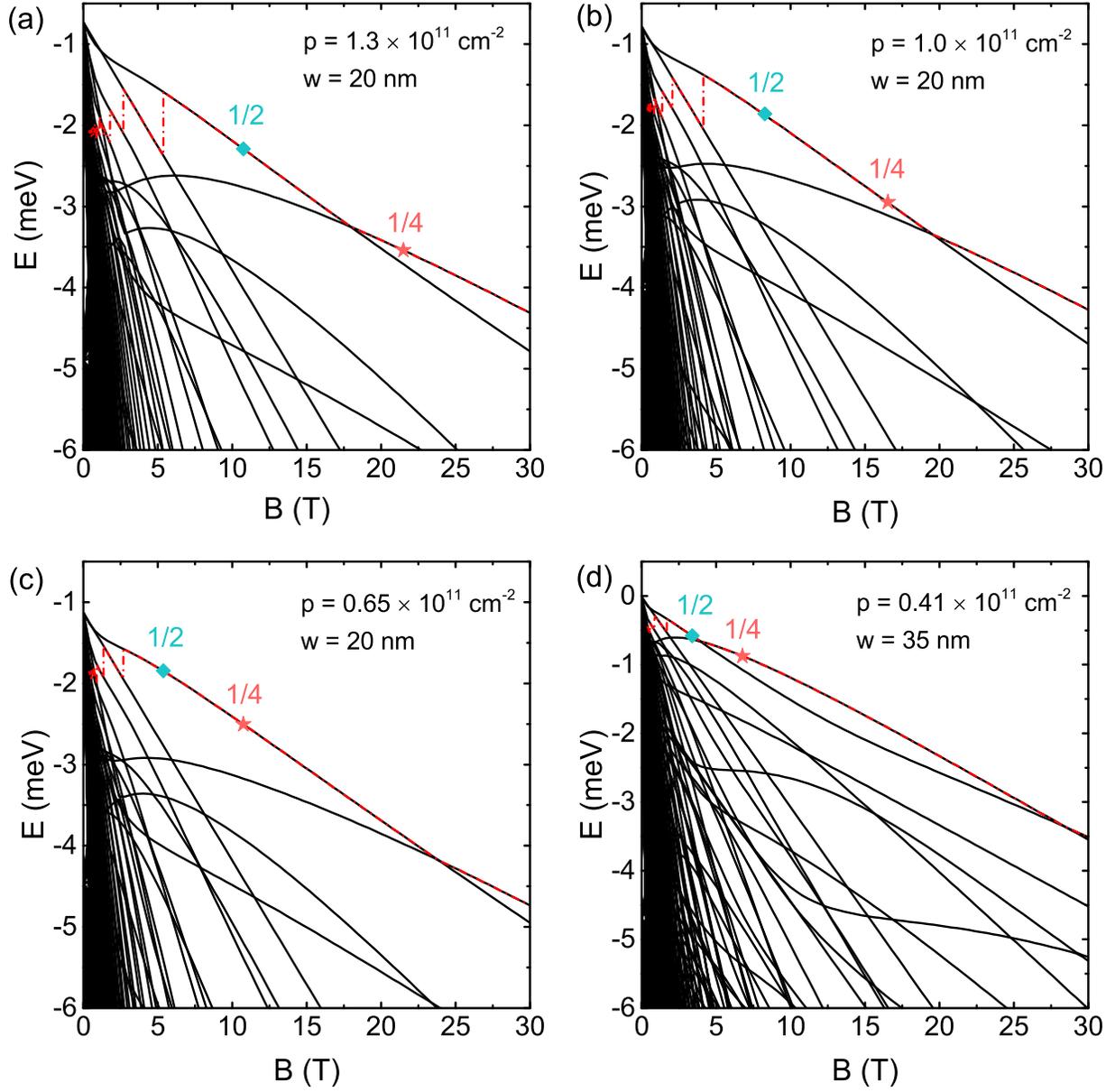

Fig. S1. Calculated energy ($E$) vs magnetic field ($B$) Landau level diagrams for our 2D hole systems with different hole densities $p$ and quantum well widths $w$: (a) $p = 1.3 \times 10^{11}$ cm$^{-2}$, $w = 20$ nm; (b) $p = 1.0 \times 10^{11}$ cm$^{-2}$, $w = 20$ nm; (c) $p = 0.65 \times 10^{11}$ cm$^{-2}$, $w = 20$ nm; (d) $p = 0.41 \times 10^{11}$ cm$^{-2}$, $w = 35$ nm. Black curves indicate different LLs, and the dash-dotted, red curve traces the position of the Fermi energy. In each plot, the field position of $\nu = 1/4$ is marked by a red star, and $\nu = 1/2$ marked by a blue diamond.

In Table I, we list the detailed parameters for the four samples we studied: quantum

TABLE I. Landau level mixing parameters ($\kappa_1$ and $\kappa_2$) for samples in our experiments

| Sample | $w$ (nm) | $x$ (%) | $p$ ($10^{11}$ cm$^{-2}$) | $m^{*\text{a}}$ ($m_0$) | $\kappa_1 = E_{Coul}/(\hbar eB/m^*)$ at $\nu = 1/2$ | at $\nu = 1/4$ | $\kappa_2 = E_{Coul}/\Delta$ at $\nu = 1/2$ | at $\nu = 1/4$ |
|---|---|---|---|---|---|---|---|---|
| A | 20 | 16 | 1.3 | 0.64 | 7.3 | 5.2 | 30 | 82 |
| B | 20 | 8 | 1.0 | 0.52 | 6.7 | 4.8 | 17 | 109 |
| C | 20 | 8 | 0.65 | 0.28 | 4.5 | 3.2 | 11 | 21 |
| D | 35 | 8 | 0.41 | 0.54 | 10.9 | 7.7 | 147 | 56 |

[a] The effective mass $m^*$ is the density-of-states effective mass obtained from band calculations at zero magnetic field.

well width ($w$), Al fraction for the Al$_x$Ga$_{1-x}$As barrier ($x$), 2D hole density ($p$), and the calculated density-of-state effective mass of holes at $B = 0$ ($m^*$). We estimate $\kappa$ at $\nu = 1/2$ and $1/4$ under the simplistic assumption of two extreme scenarios:

(i) Assume a constant $m^*$ and linear LLs as a function of $B$. In this case, $\kappa_1$, defined as the ratio between $E_{Coul}$ and "cyclotron energy" ($\hbar eB/m^*$), is proportional to $1/\sqrt{B}$. Therefore $\kappa_1$ at $\nu = 1/2$ is always a factor of $\sqrt{2}$ larger than $\kappa_1$ at $\nu = 1/4$ because the $B$ position of $\nu = 1/2$ is half of the $\nu = 1/4$ $B$ position. The values of $\kappa_1$ at $\nu = 1/2$ and $1/4$ are listed in Table I. However, this is not a very good approximation for GaAs 2DHSs, especially at high $B$ where LLs are highly nonlinear and exhibit multiple crossings, as shown in Fig. S1. We also note that the slopes of LLs at large $B$ are generally smaller than near zero field, suggesting that $\kappa_1$ is an underestimate of $\kappa$ at large $B$.

(ii) Assume $E_{cyc}$ is given by the separation $\Delta$ between the two lowest-energy LLs. In this case, $\kappa_2$, defined as the ratio between $E_{Coul}$ and $\Delta$, is strongly affected by the non-linearity of the hole LLs. The values of $\kappa_2$ at $\nu = 1/2$ and $1/4$ are listed in Table I. These values are overall much larger than those of $\kappa_1$. It is worth noting that the exchange interaction, as well as the mixing between hole LLs with opposite parities, can be significantly suppressed [2]. Therefore $\kappa_2$ is an overestimate of the true $\kappa$.

We emphasize that although the calculated LLs are known to capture the main features for LLs in GaAs 2DHSs (e.g., nonlinearity and existence of crossings between LLs), there can be quantitative discrepancies between the predicted the experimentally observed $B$ positions of the crossings [3, 4]. The values of $\kappa_1$ and $\kappa_2$ obtained from the $E$ vs $B$ LL diagrams should

therefore be treated cautiously.

## II. HALL DATA

Figure S2 shows Hall traces for sample A taken at $T \simeq 120$ mK for opposite polarities of magnetic field ($B^+$ in black, $B^-$ in blue). Note that the $|R_{xy}|$ traces for $B^+$ and $B^-$ are close between 18 and 20.3 T, and exhibit quantized Hall plateaus at fractional fillings $\nu = 2/7$, 3/11, and 4/15. At $B > 20.3$ T, however, $B^+$ and $B^-$ traces significantly deviate from the linear trend because of the significant $R_{xx}$ mixing when the 2DHS enters a highly insulating regime. In principle, the effect of $R_{xx}$ mixing can be minimized by averaging $|R_{xy}|$ taken for opposite polarities of $B$ [5, 6]. We did such averaging and, it successfully removes the $R_{xx}$ mixing between 18 and 20.3 T when the $R_{xx}$ values are reasonably small ($\sim$ a few k$\Omega$) but fails at higher field when $R_{xx} \sim$ M$\Omega$. The extremely high values of $R_{xx}$ and the ensuing large $R_{xx}$ mixing in $R_{xy}$ measurement at $\nu = 1/4$ and the nearby fillings preclude us from measuring $R_{xy}$ accurately and demonstrating its quantization at $4h/e^2$. This is a well-known

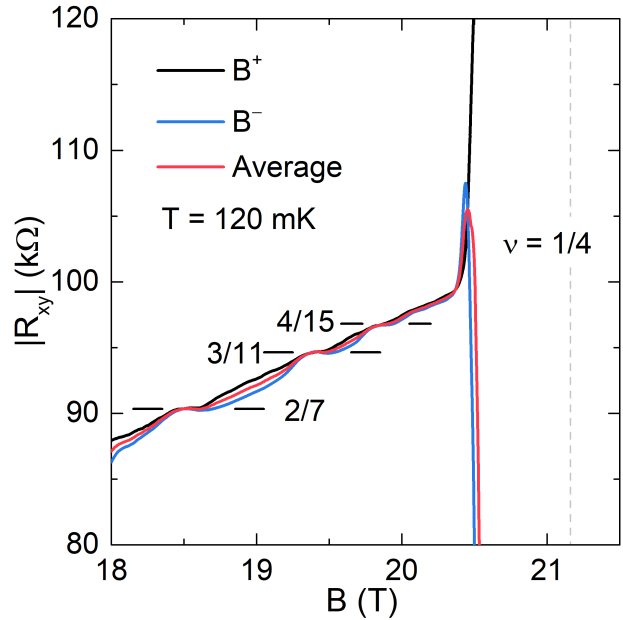

Fig. S2. Hall traces for sample A taken at $T \simeq 120$ mK for two opposite polarities of magnetic field ($B^+$ in black, $B^-$ in blue). The red trace shows the average of the $B^+$ and $B^-$ traces. Quantized Hall plateaus are observed at $\nu = 2/7$, 3/11, and 4/15.

problem and manifests itself, e.g., for the developing FQHS at $\nu = 1/7$ in 2DESs [7].

## III. PARTICLE-HOLE SYMMETRY

Assuming that particle-hole symmetry holds in the lowest LL in our GaAs 2DHSs, the $\nu = 1/4$ and $3/4$ FQHSs we observe may be viewed as a pair of particle-hole conjugate states. In order to explore the possible connection between the correlated states near $\nu = 1/4$ and $3/4$, in Fig. S3 we present $R_{xx}$ vs $\nu$ traces near $\nu = 1/4$ (top panels), and $R_{xx}$ vs $(1 - \nu)$ traces near $\nu = 3/4$ (bottom panels) for samples A and B. The traces near $\nu = 1/4$ are measured with an excitation current of 0.1 nA, and at $T \simeq 95$ and 100 mK for sample A and B, respectively. The traces near $\nu = 3/4$ are measured at $T \simeq 20$ mK with an excitation current of 20 nA. The traces near $\nu = 1/4$ behave qualitatively different from those near $\nu = 3/4$ for both samples: (i) Although a local $R_{xx}$ minimum is observed at both $\nu = 1/4$ and $3/4$, $R_{xx}$ near $\nu = 1/4$ is extremely large ($\sim 10$ M$\Omega$), while $R_{xx}$ near $\nu = 3/4$ is only $\sim 1$

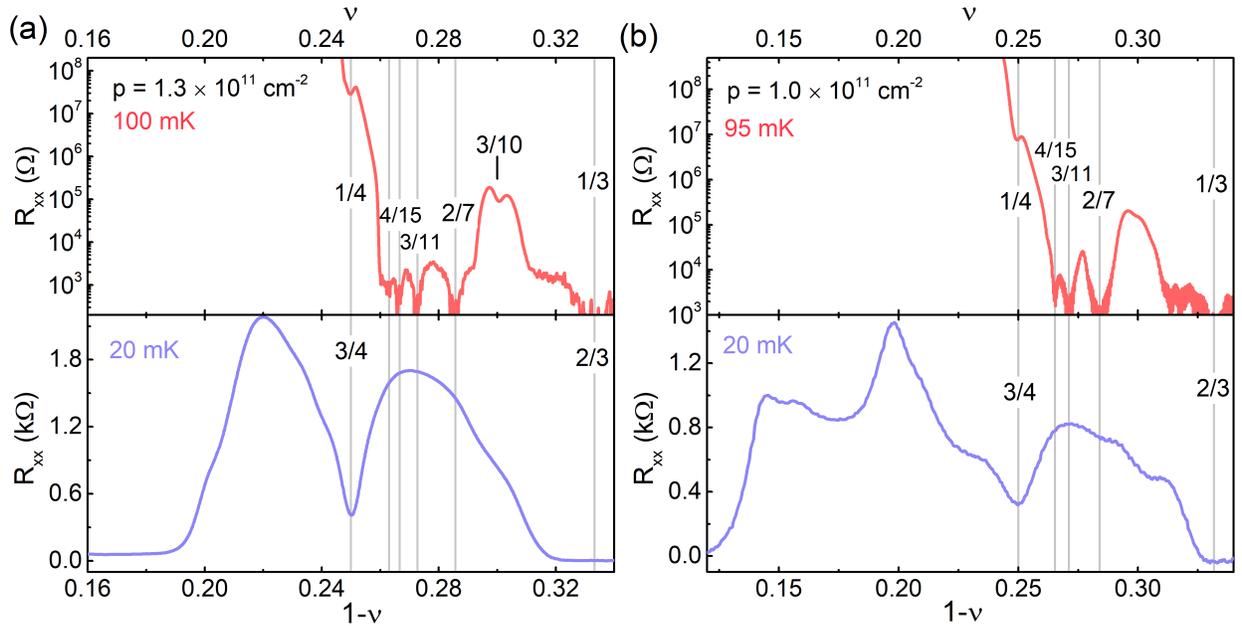

Fig. S3. Magneto-resistance $R_{xx}$ near $\nu = 1/4$ and $3/4$ for 2DHSs with densities (a) $p = 1.3 \times 10^{11}$ cm$^{-2}$, and (b) $p = 1.0 \times 10^{11}$ cm$^{-2}$. In order to compare the data at $\nu$ and $(1 - \nu)$, the traces in the top panel are plotted vs $\nu$, and the bottom panel vs $(1 - \nu)$, so that $\nu$ near $1/4$ is aligned with $(1 - \nu)$ near $3/4$. Several pairs of $\nu$ and $(1 - \nu)$ are marked by grey solid lines.

kΩ. (ii) In the top panels, numerous $R_{xx}$ minima are observed at odd-denominator fillings in the range $1/4 < \nu < 1/3$, at $\nu = 2/7$, $3/11$, and $4/15$. In the bottom panels, however, $R_{xx}$ minima are absent at their particle-hole conjugate fillings in the range $2/3 < \nu < 3/4$.

## IV. TILTED MAGNETIC FIELD DATA

We studied the evolution of $R_{xx}$ traces with the application of an additional in-plane magnetic field ($B_\parallel$), which we introduce by tilting the 2DHS *in situ* (Fig. S4 inset). Such data shed light on an important characteristic of the $\nu = 1/4$ FQHS we observe, namely its spin polarization; this is important because a requirement for a FQHS, such as the one seen at $\nu = 5/2$, to be non-Abelian is that it is fully spin polarized [8, 9]. The data, shown in Fig. S4 for sample D, reveal that $R_{xx}$ background in the low filling regime becomes larger as we tilt the sample, but the $R_{xx}$ minimum at $\nu = 1/4$ is always present. This insensitivity of the $\nu = 1/4$ $R_{xx}$ minimum to $B_\parallel$, persists up to $B_\parallel \simeq 12$ T (total magnetic field $B \simeq 14$

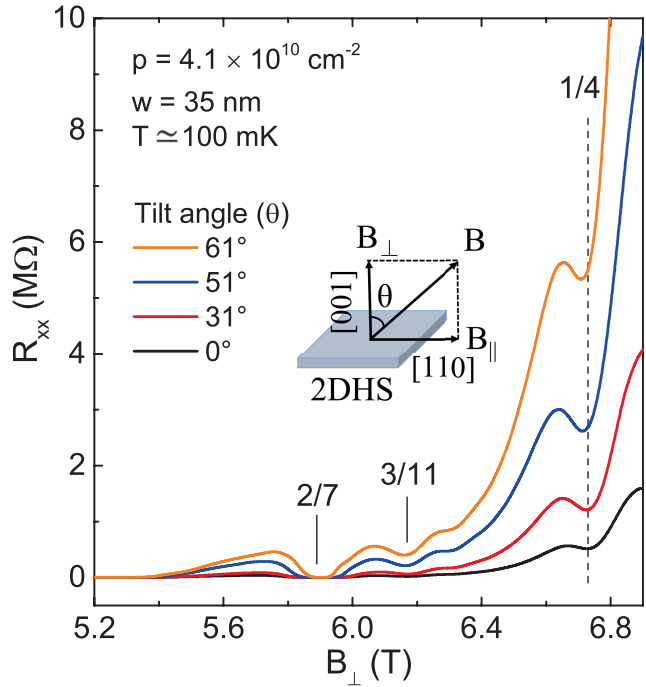

Fig. S4. $R_{xx}$ vs $B_\perp$ traces for sample D taken at different tilt angles. Inset shows a schematic of the experimental setup for applying tilted fields; the sample is mounted on a rotating stage to support *in situ* tilt.

T). Such a large $B$ should make the 2DHS fully spin polarized. The absence of a spin transition, therefore, implies that the $\nu = 1/4$ FQHS in our 2DHS is fully spin polarized. This conclusion is also supported by our observation of the $\nu = 1/4$ FQHS in a wide range of perpendicular magnetic fields ($7 \lesssim B_\perp \lesssim 21$ T) in samples with different densities.

Fully-Spin-Polarized Fermi Sea near $\nu = 5/2$, Phys. Rev. Lett. **120**, 256601 (2018), and the references therein.



\end{document}